\documentclass[cits]{PoS}
\usepackage{multicol}
\usepackage{multirow}
\usepackage{slashed}
\usepackage{setspace}
\usepackage{color}
\usepackage{booktabs}
\usepackage{graphicx}
\DeclareMathAlphabet{\mathcal}{OMS}{cmsy}{m}{n} 

\title{Neutral B-meson mixing parameters in and beyond the SM with $2+1$ flavor lattice QCD}

\ShortTitle{Neutral B-meson mixing parameters}

\author{
C.M.~Bouchard\textsuperscript{\textnormal{\textit{a,b}}}, 
E.D.~Freeland\textsuperscript{\textnormal{\textit{c}}}, 
C.W.~Bernard\textsuperscript{\textnormal{\textit{d}}}, 
C.C.~Chang\textsuperscript{\textnormal{\textit{e,f}}}, 
{\speaker{A.X.~El-Khadra}\textsuperscript{\textnormal{\textit{ e}}}}, 
M.E.~G\'amiz\textsuperscript{\textnormal{\textit{g}}}, 
A.S.~Kronfeld\textsuperscript{\textnormal{\textit{f,h}}}, 
J.~Laiho\textsuperscript{\textnormal{\textit{i}}}, 
R.S.~Van de Water\textsuperscript{\textnormal{\textit{f}}}\\
    \textsuperscript{a}Department of Physics, The Ohio State University, Columbus, OH 43210, USA\\
    \textsuperscript{b}Department of Physics, The College of William and Mary, Williamsburg, VA  23187, USA\\
        \textsuperscript{c}Liberal Arts Department, The School of the Art Institute of Chicago, Chicago, 
        IL 60603, USA\\
    \textsuperscript{d}Department of Physics, Washington University, St. Louis, MO 63130, USA\\
    \textsuperscript{e}Physics Department, University of Illinois, Urbana, IL 61801, USA\\
    \textsuperscript{f}Theoretical Physics Department, Fermi National Accelerator Laboratory,\thanks{Operated
        by Fermi Research Alliance, LLC, under Contract No.\ DE-AC02-07CH11359 with the United 
        States Department of Energy} { } Batavia, IL 60510, USA\\ 
    \textsuperscript{g}CAFPE and Departamento de Fisica Teorica y del Cosmos, Universidad de Granada, 
        E-18002 Granada, Spain\\
    \textsuperscript{h}Institute for Advanced Study, Technische Universit\"at M\"unchen,
        85748 Garching, Germany\\
    \textsuperscript{i}Department of Physics, Syracuse University, Syracuse, NY 13244, USA\\}
\author{Fermilab Lattice and MILC Collaborations\\
        E-mail: \email{axk@illinois.edu}}

\abstract{We report on the status of our calculation of the hadronic matrix elements for neutral $B$-meson mixing 
with asqtad sea and valence 
light quarks and using the Wilson clover action with the Fermilab interpretation for the $b$ quark. We calculate the 
matrix elements of all 
five local operators that contribute to neutral $B$-meson mixing both in and beyond the Standard Model. We use MILC 
ensembles with $N_f=2+1$ 
dynamical flavors at four different lattice spacings in the range $a \approx 0.045$--$0.12$~fm, and with light 
sea-quark masses as low as 
0.05 times the physical strange quark mass. We perform a combined chiral-continuum extrapolation 
including the so-called wrong-spin 
contributions in simultaneous fits to the matrix elements of the five operators. We present a complete 
systematic error budget and 
conclude with an outlook for obtaining final results from this analysis.}

\FullConference{The 32nd International Symposium on Lattice Field Theory\\
		 23-28 June, 2014\\
		 Columbia University New York, NY}

\begin{document}

\section{Introduction and motivation}

Neutral meson mixing, which is loop-induced in the Standard Model (SM), plays an important role in determining the 
CP violating parameters of the SM as  well as in providing constraints on BSM theories. 
In the SM, neutral $B$-meson mixing receives contributions from hadronic matrix elements of 
$\Delta B = 2$ local operators. In BSM theories, additional
$\Delta B = 2$ local operators can contribute, and the most general $\Delta B = 2$ effective hamiltonian can be written
in terms of five  operators,
\begin{equation}
{\cal H}_{\rm eff} = \sum_{i=1}^5 C_i \, {\cal O}_i \, ,
\end{equation}
where the integrated-out high-momentum physics is collected into the Wilson coefficients, $C_i$, which 
therefore depend on the underlying theory (SM or BSM). In a commonly used basis the local $\Delta B=2$ 
operators ${\cal O}_i$ take the form
\begin{equation}
 {\cal O}_1 =
   \left( \bar{b}\gamma_\mu \,L \, q \right)
   \left( \bar{b}\gamma_\mu \, L \, q \right)\, , \qquad
 \mathcal{O}_2 =  \left( \bar{b}\, L \, q \right)
   \left( \bar{b} \, L \, q \right)   \, ,
   \label{eq:Q1}
\end{equation}
\begin{displaymath}
\mathcal{O}_3 =  \left( \bar{b}^{\alpha} L \, q^{\beta} \right)
   \left( \bar{b}^{\beta} L \, q^{\alpha} \right) \, , \quad
\mathcal{O}_4 =  \left( \bar{b} \, L \, q \right)
   \left( \bar{b}\, R \, q\right) \, , \quad
\mathcal{O}_5 =  \left( \bar{b}^{\alpha}  L \, q^{\beta} \right)
   \left( \bar{b}^{\beta} R \, q^{\alpha} \right)\, ,
\end{displaymath}
where $q=d,s$ denotes a light (down or strange) quark,
and $R,L = \frac{1}{2} (1 \pm \gamma_5)$. 
The superscripts $\alpha,\beta$ are color indices, which
are shown only when they are contracted across the two bilinears. 
The SM prediction for the neutral $B_q$-meson ($q=d,s$) mass difference is given by
\begin{equation}
\Delta M_q = \left( \frac{G_F^2 M_W^2 S_0}{4 \pi^2 M_{B_q}} \eta_B(\mu) \right)  |V_{tq}^* V_{tb}|^2 \langle  \mathcal{O}_1  \rangle (\mu),
\label{eq:Bmix}
\end{equation}
where the quantities in parentheses are known and include short-distance QCD and EW corrections; $M_{B_q}$ is the mass of the
$B_q$ meson; and,  $\langle  \mathcal{O}_i  \rangle (\mu) \equiv \langle  \bar{B}^0_q | \mathcal{O}_i | B^0_q \rangle (\mu)$ is the hadronic 
matrix element of $\mathcal{O}_i$, which can be parameterized in terms of the decay constant $f_{B_q}$ and bag parameter 
$B_{B_q}$ (see, for example, Section 8 of Ref.~\cite{Aoki:2013ldr} for further details).  The mass differences are measured
to sub-percent accuracy \cite{Agashe:2014kda}, and the determination of the CKM parameters in Eq.~(\ref{eq:Bmix}) is limited by the 
theory uncertainty on the hadronic matrix elements. 
The ratio of the $B_s$ and $B_d$  mass differences is of particular interest in unitarity triangle analyses, due to the 
cancellation of statistical and several systematic errors in the corresponding ratio of matrix elements:
\begin{equation}
\frac{\Delta M_s}{\Delta M_d} \, \frac{M_{B_d}}{M_{B_s}} = \left| \frac{V_{ts}}{V_{td}} \right|^2 \, \xi^2 \qquad {\rm with} 
\quad \xi^2 \equiv \frac{f^2_{B_s} \hat{B}_{B_s}}{f^2_{B_d} \hat{B}_{B_d}} \,.
\end{equation}
There are several published lattice QCD calculations of $\xi$ with $N_f=2+1$ \cite{Gamiz:2009ku,Albertus:2010nm,Bazavov:2012zs}
and $N_f=2$  \cite{Carrasco:2013zta} flavors of sea quarks, and the current FLAG review \cite{Aoki:2013ldr} quotes an uncertainty 
of $5\%$ for $\xi$. Most of the previous lattice calculations have focused on the matrix elements of $\mathcal{O}_{1-3}$ needed for
the SM predictions of the mass and width differences. For the matrix elements of 
all five operators, results with $N_f=2$ dynamical flavors have been reported \cite{Carrasco:2013zta}.  
Ours is the first lattice QCD calculation of all five matrix elements with $N_f=2+1$ dynamical flavors. 
At this conference, preliminary results for $\langle\mathcal{O}_{1,2,3}\rangle$ with $N_f=2+1+1$ flavors with 
physical light quarks were presented~\cite{Dowdall:2014qka}.

\section{Lattice set-up}

In this work we use the fourteen MILC asqtad ensembles \cite{Bazavov:2009bb} with $N_f=2+1$ dynamical flavors of sea quarks listed in 
Table~1 of Ref.~\cite{Chang:2013gla}. 
Included are ensembles at four different lattice spacings covering the range $a \approx 0.045$--$0.12$~fm. At every lattice spacing
(except for the finest) we have four or five ensembles with different light sea-quark masses, the lowest of which has a mass of about 
0.05 times the physical strange quark mass. The light valence quarks also employ the asqtad action, where on each sea-mass 
ensemble we generate propagators with at least seven valence quark masses covering the range from the physical strange quark 
mass to the lightest sea quark mass. 
The $b$-quark propagators are generated with the Wilson clover action with the Fermilab interpretation \cite{ElKhadra:1996mp}, where the hopping 
parameter $\kappa_b$ is tuned to give the experimental result for
the $B_s$ meson mass. The $b$-quark fields in the four-quark operators are rotated  so that 
the operators are $O(a)$ improved \cite{Bazavov:2012zs}, the same as the heavy-quark action.

Compared to our previously reported results for $\xi$ \cite{Bazavov:2012zs}, 
in our present analysis we 
have more than double the number
of ensembles, including two additional finer lattice spacings, smaller light-quark masses, and a significant increase in statistics. 
Our previously reported preliminary results for all five matrix elements \cite{Bouchard:2011xj} were also obtained on a smaller subset of MILC
ensembles than in the present analysis.  The current analysis adds the finest lattice spacing and a few ensembles with lighter sea quark masses at the other lattice spacings. 

On each ensemble and for each valence quark mass, we generate the two- and three-point functions needed to extract the 
hadronic matrix elements. This analysis step was described previously \cite{Bazavov:2012zs,Bouchard:2011xj,Freeland:2012kz},
and our 
results for the matrix 
elements from the correlator fits were presented in Ref.~\cite{Chang:2013gla}. We obtain renormalized matrix elements in the 
$\mathrm{\overline{MS}}$-NDR scheme using one-loop mean field improved
perturbation theory \cite{Bazavov:2012zs,Bouchard:2011xj}, which yields:
\begin{equation}
\langle \mathcal{O}_i \rangle^{\mathrm{\overline{MS}-NDR}}(m_b)=\sum_{j=1}^5[ \delta_{ij}+\alpha_s \zeta_{ij} ] \langle \mathcal{O}_j \rangle^{\rm lat}.
\label{eq:renorm}
\end{equation}
The $\zeta_{ij}$ are the differences between the continuum and lattice renormalizations, and the coupling $\alpha_s$ is 
evaluated as described
in Refs.~\cite{Bazavov:2012zs,Bouchard:2011xj}. For our final results we will quote the matrix elements for both the 
BBGLN \cite{Beneke:1998sy} and BJU \cite{Buras:2001ra} choices for 
the evanescent operators.  
  
The last step before the chiral-continuum extrapolation is to correct the matrix elements for mistunings of the heavy-quark mass parameter 
$\kappa_b$. With better statistics, we find that 
our current best estimates of this parameter
(see Ref.~\cite{Bailey:2014tva}) differ from  
the simulation values for $\kappa_b$. 
Using $r_1$ units, we correct the matrix 
elements by applying a linear shift with respect to the inverse kinetic meson mass, $M_2^{-1}$ 
(where we omit factors of $r_1$ for simplicity): 
\begin{equation}
\langle \mathcal{O}_i \rangle_{\rm corrected} = \langle \mathcal{O}_i \rangle + \frac{ \partial  \langle \mathcal{O}_i \rangle}{\partial M_2^{-1}} \Delta M_2^{-1}. 
\end{equation} 
The slope 
$\partial  \langle \mathcal{O}_i \rangle / \partial M_2^{-1}$ is obtained from the heavy-quark mass dependence of the 
$\langle \mathcal{O}_i \rangle$ calculated on one ensemble, and  $\Delta M_2^{-1}$ is the difference in
the $B$-meson $M_2$ due to  changing $\kappa_b$ from its simulation value to its tuned value. 
The error associated with this correction (slope and tuned $\kappa_b$) is accounted for via Bayesian constrained coefficients in the chiral-continuum extrapolation and is therefore included in our statistical fit error. 

\section{Chiral-continuum extrapolation}

We use SU(3), heavy-meson, rooted, staggered, partially quenched, chiral perturbation theory 
to perform 
a combined chiral and continuum extrapolation of the $\langle \mathcal{O}_i\rangle$ \cite{Bernard:2013dfa}. For example,
at NLO in $\chi$PT  the expansion for $\langle \mathcal{O}_1 \rangle$ takes the form:  
\begin{eqnarray}
\langle \mathcal{O}_1\rangle_q & = & \beta_1 \left( 1+ \frac{W_{q\bar{b}}+W_{b\bar{q}}}{2} 
 + T_q + Q_q + \tilde{T}_q^{(a)} + \tilde{Q}_q^{(a)} \right)
 + 2 \, (\beta_2 + \beta_3 )  \tilde{T}_q^{(b)} 
 + 2 \, (\beta'_2 + \beta'_3 )  \tilde{Q}_q^{(b)}   \nonumber \\
 & & + {\rm \, analytic \; terms}. 
\label{eq:chpt}
\end{eqnarray}
where the $W,T,Q$'s are the usual chiral logarithms modified to include taste-breaking effects \cite{Aubin:2005aq}. 
The $\tilde{T}, \tilde{Q}$ terms are additional taste-changing effects that contribute at NLO due to our choice of
using a local staggered field in the four-quark operators. With simultaneous fits to 
$\left[ \langle \mathcal{O}_1 \rangle, \langle \mathcal{O}_2 \rangle,\langle \mathcal{O}_3 \rangle \right]$ and 
$[\langle \mathcal{O}_4 \rangle, \langle \mathcal{O}_5 \rangle]$, 
respectively, no new fit parameters are required, because the additional terms are proportional to the LECs
$\beta_i$ and~$\beta'_i$. The leading corrections to the heavy-meson expansion arise at $O(1/M)$ and are included in our fits via 
the hyperfine and flavor splittings. We vary the chiral-continuum fit function to study the systematics associated with the 
combined extrapolation by adding higher-order analytic terms with Bayesian constrained coefficients, where the constraints are
guided by power-counting expectations. We find that the inclusion of NNLO analytic terms results in stable fits, where the 
central values and errors don't change appreciably if we add additional terms to our fit function. 

\begin{figure}
    \centering
    \includegraphics[width=0.421\textwidth,angle=270]{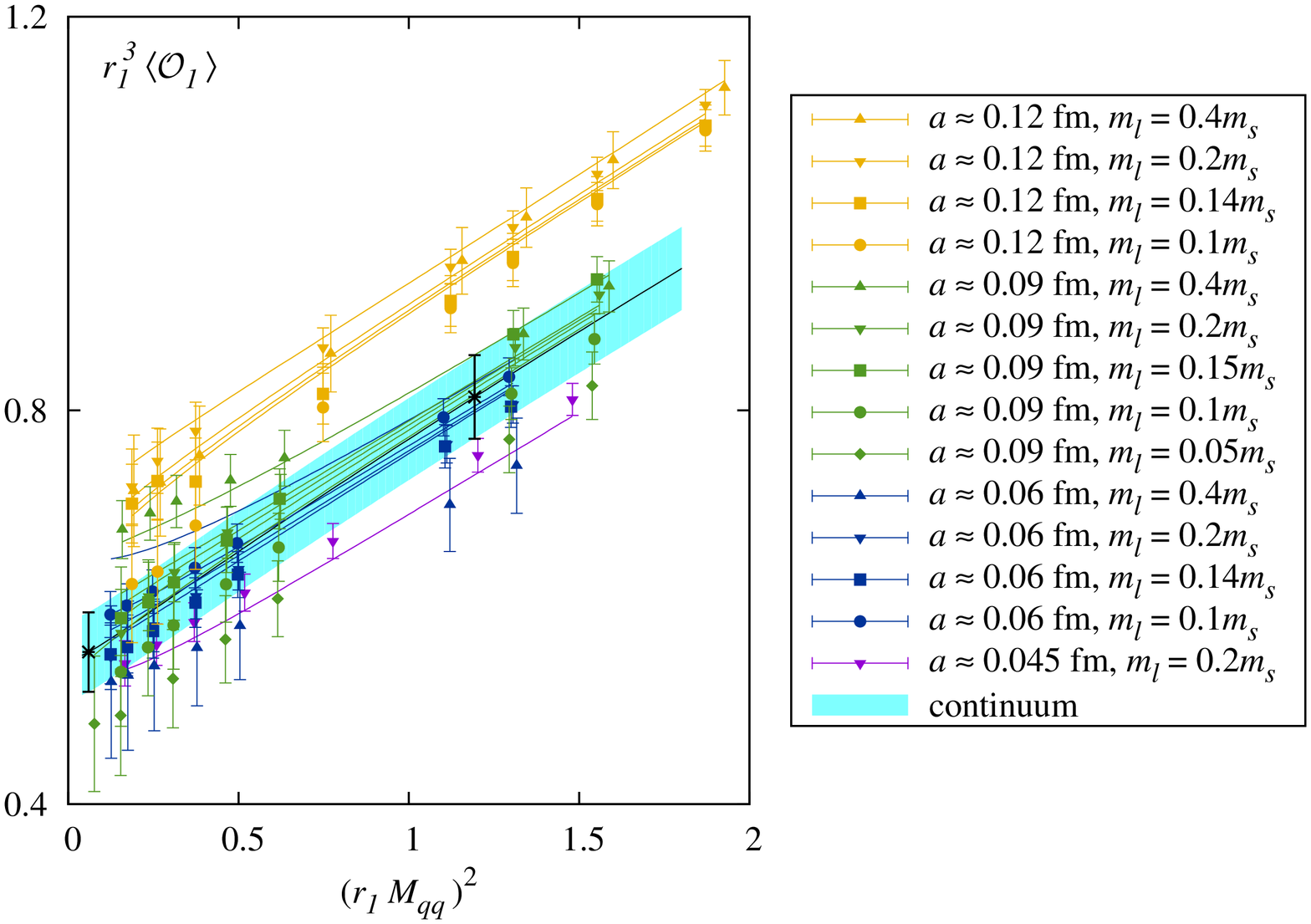}
    \includegraphics[width=0.421\textwidth,angle=270]{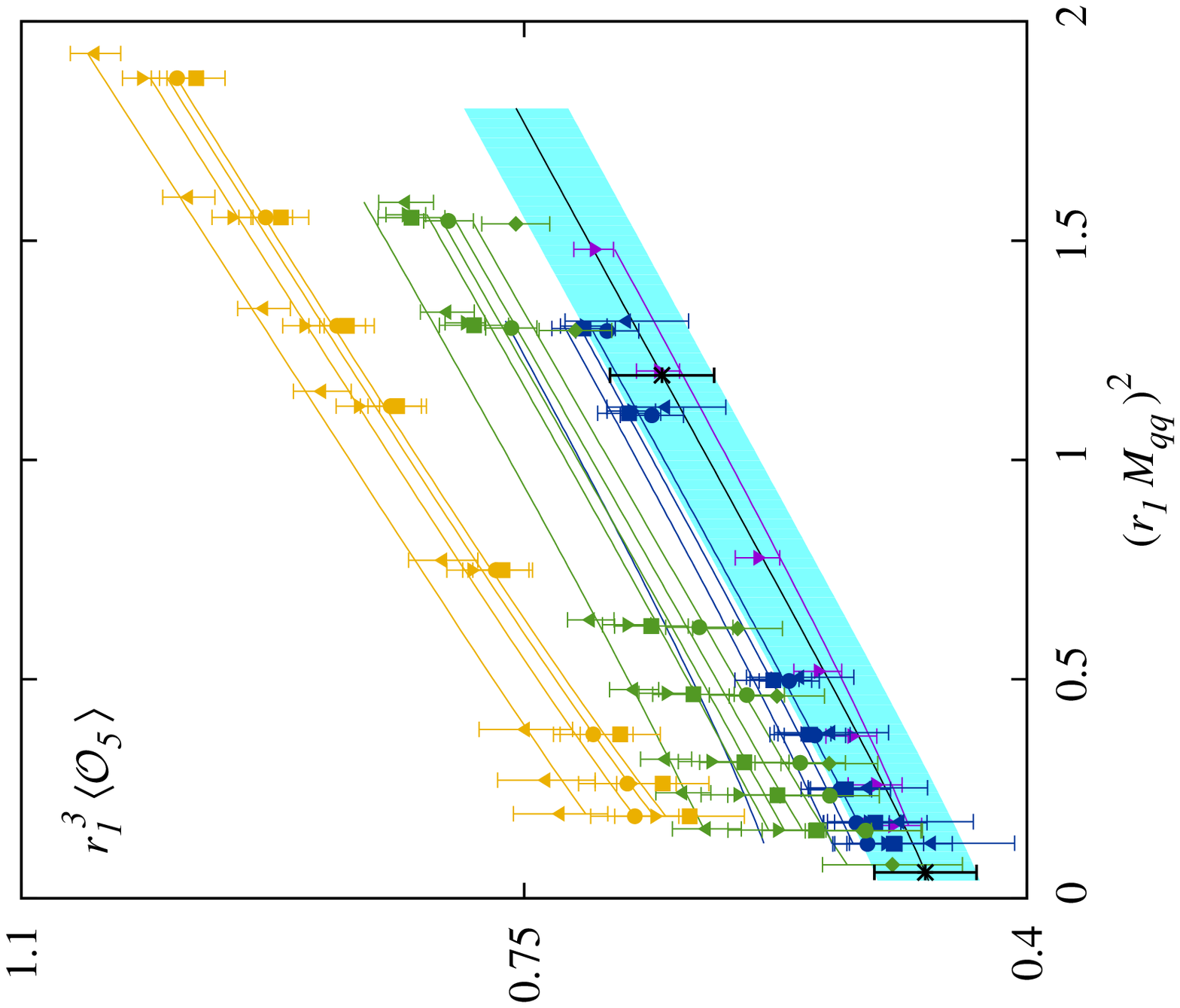}
    \vspace{-0.5cm}
    \caption{Chiral-continuum extrapolation of the matrix elements $\langle \mathcal{O}_1 \rangle$ (left panel) and $\langle \mathcal{O}_5 \rangle$ (right panel) in $r_1$ units. 
    The data points are the renormalized matrix elements shown as functions of the valence taste-pion mass-squared for the different lattice spacings and 
    sea quark masses, as indicated in the legend box. The lines indicate the result of  the fit, with the black line and cyan band showing the continuum 
    extrapolation.} 
    \label{fig:O1O5extrap}
\end{figure}

Generic light-quark discretization terms of $O(\alpha_s a^2, a^4)$ and taste-violating terms of $O(\alpha^2_s a^2)$ 
are included in the analytic terms in Eq.~(\ref{eq:chpt}).  
We also add heavy-quark discretization effects to our fit function. With the Fermilab interpretation discretization effects 
arise due to a mismatch between the coefficients of the lattice and continuum HQETs and result in mass-dependent coefficients.
 Heavy-quark discretization errors then take the form $\sim f_k (am_b)(a \Lambda)^n$.  
We include heavy-quark discretization terms of $O(\alpha_s a, a^2, a^3)$ in our fit function, where we chose $\Lambda=800$~MeV. 
Figure~\ref{fig:O1O5extrap} shows sample chiral-continuum fits for $\langle \mathcal{O}_1 \rangle$ and $\langle \mathcal{O}_5 \rangle$. 

\section{Systematic error budget}

\begin{figure}
    \centering
    \includegraphics[width=0.29\textwidth,angle=270]{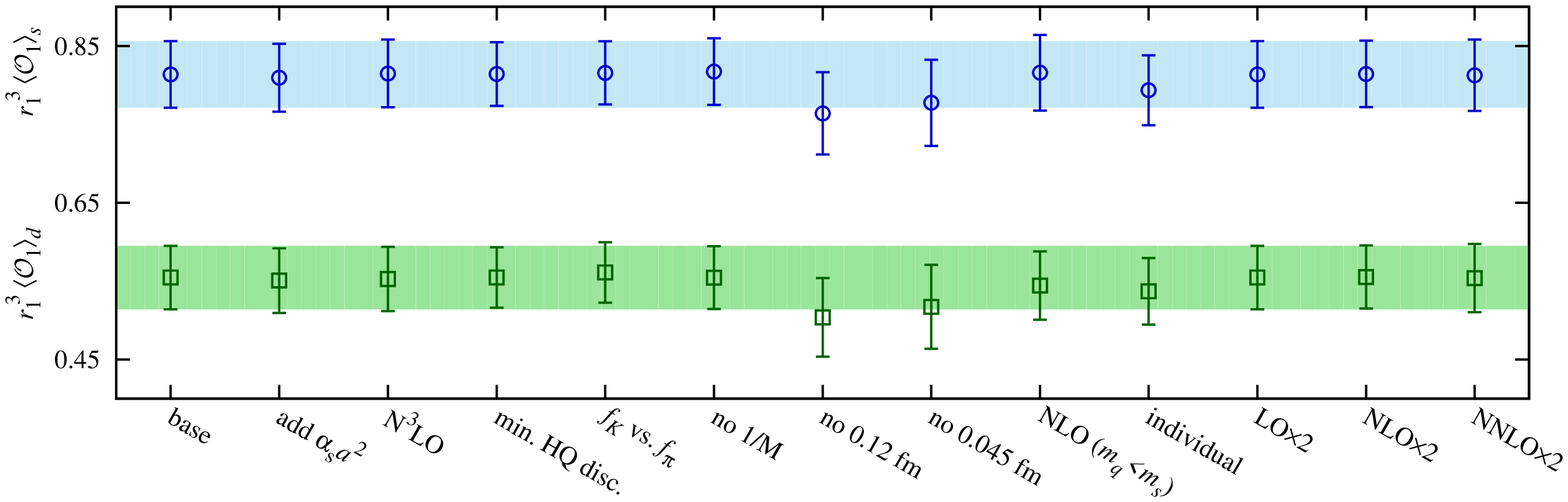}
     \includegraphics[width=0.29\textwidth,angle=270]{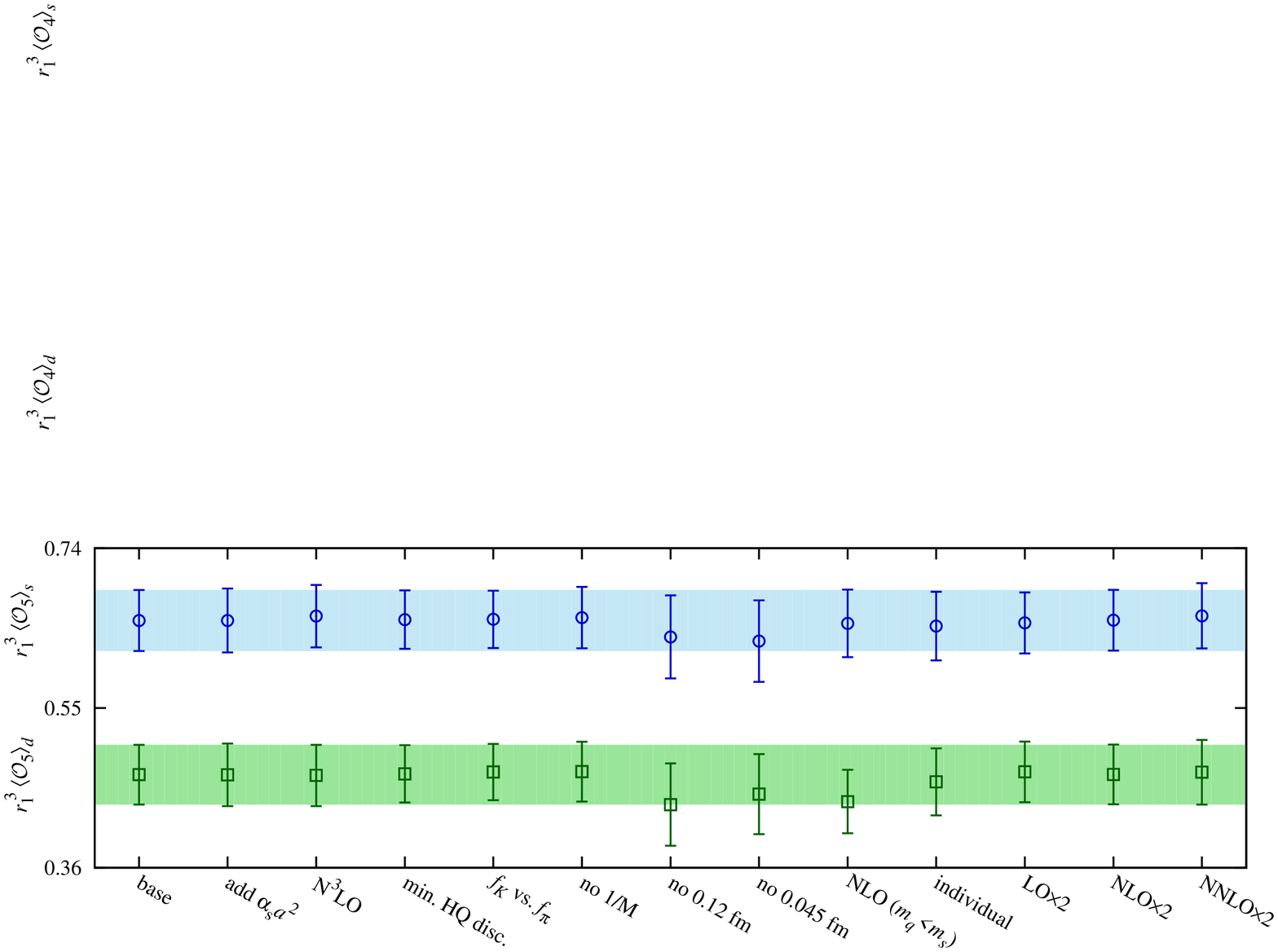}
     \vspace{-0.2cm}
    \caption{Stability plots for $\langle \mathcal{O}_1 \rangle$ and $\langle \mathcal{O}_5 \rangle$ showing the variation of the chiral-continuum fits with different choices for the fit function and parameters.
    The bands show the result of the preferred fits.} 
    \label{fig:stability}
   \vspace{-0.5cm}
\end{figure}

The dominant systematic errors in our calculation are due to the chiral-continuum extrapolation, heavy-quark discretization effects, and the perturbative matching
of the four-quark operators. In the two former cases, we account for the error due the truncation of the corresponding expansions by considering 
fits that include more (higher order) terms with Bayesian constrained coefficients until the results (central values and error bars) stabilize. In this way, 
the statistical fit error includes the systematic error from truncation. This is illustrated in 
Figure~\ref{fig:stability} for $\langle \mathcal{O}_1 \rangle$ and $\langle \mathcal{O}_5 \rangle$. The stability plots for the other matrix elements are very similar.  
We see that our fits are stable under adding $O(\alpha_s a^2)$ terms, adding N$^3$LO analytic terms, 
dropping higher-order HQ discretization terms, using $f_K$ instead of $f_\pi$, dropping the $1/M$ terms from the heavy-meson expansion, excluding
data from the coarsest (finest) lattice spacing, and increasing the prior widths. 

\renewcommand*\arraystretch{1.2}
\begin{table}
\begin{center}
\begin{tabular}{l cc cc cc cc }
 								& 				& 								& 		& 							& 			&  	& \vspace{-0.15in} 			\\
 		& \multicolumn{2}{c}{$\langle \mathcal{O}_1 \rangle_s$} & \multicolumn{2}{c}{$\langle \mathcal{O}_1 \rangle_d$} & 
		\multicolumn{2}{c}{$\langle \mathcal{O}_i \rangle_s$ ( $i>1$)}   &  \multicolumn{2}{c}{$\langle \mathcal{O}_i \rangle_d$ ($i>1$)}  \\
 								& 				& 								& 		& 							& 			& 	&  & \vspace{-0.15in} 	\\	 \cline{1-9} \\ 
 								& 				& 								& 		& 							& 			& 	& & \vspace{-0.425in}	\\ \cline{1-9}
 \multicolumn{1}{r|}{source}			& 2011			& \multicolumn{1}{c|}{2014}  			& 2011	& \multicolumn{1}{c|}{2014} 		& 2011		& \multicolumn{1}{c|}{2014} & 2011		& 2014 		\\  \cline{1-9}
 \multicolumn{1}{r|}{} 					&  		 		& \multicolumn{1}{c|}{} 				& 		& \multicolumn{1}{c|}{} 	 		& 			&   \multicolumn{1}{c|}{} 	&  &  \vspace{-0.15in} 	\\
 \multicolumn{1}{r|}{comb. stat. $\chi$PT}			& 7			& \multicolumn{1}{c|}{5} 		& 15 &  \multicolumn{1}{c|}{7}			& 3--11		&  \multicolumn{1}{c|}{4--12} & 	4.3--16		& 6--15 \\  \cline{1-9}
 \multicolumn{1}{r|}{HQ disc.}			& 4				& \multicolumn{1}{c|}{included}	 		& 4		& \multicolumn{1}{c|}{included}		& 4			& \multicolumn{1}{c|}{included}	 & 4			& included			\\  \cline{1-9}
 \multicolumn{1}{r|}{inputs}			& 5.1			& \multicolumn{1}{c|}{included}	 		& 5.1		& \multicolumn{1}{c|}{included}		& 5.1			&  \multicolumn{1}{c|}{included}  &	5.1			& included	\\  \cline{1-9}
 \multicolumn{1}{r|}{renormalization}				& 8			& \multicolumn{1}{c|}{6.4}	  			& 8		& \multicolumn{1}{c|}{6.4}	 		&  8	& \multicolumn{1}{c|}{6.4}	 		& 8		& 6.4		\\  \cline{1-9}
 \multicolumn{1}{r|}{finite volume}				& 1			& \multicolumn{1}{c|}{1}	  			& 1		& \multicolumn{1}{c|}{1}			& 1			& \multicolumn{1}{c|}{1}		&1	& 1 		\\ \cline{1-9}
 \multicolumn{1}{r|}{total}	& 	12		& \multicolumn{1}{c|}{8} 	& 18		& \multicolumn{1}{c|}{10}			& 10--15		&  \multicolumn{1}{c|}{8--13}	& 11--19  & 9--17	\\ \cline{1-9} 
 	& 				& 								& 		& 							& 			&  &	& \vspace{-0.195in}	\\ \cline{1-9}
\end{tabular}
\end{center}
\caption{Comparison of the error budgets for $\langle \mathcal{O}_i \rangle$ from Ref.~\cite{Bouchard:2011xj} (2011) with this analysis (2014).
Here, ``included'' means these errors are now included via Bayesian priors in the combined statistical, $\chi$PT error.}
\label{tab-error}
\end{table}

\renewcommand*\arraystretch{1.2}
\begin{table}
\begin{center}
\begin{tabular}{l cc }
 								& 				& 	\vspace{-0.15in} 	\\	 \cline{1-3} 
								& 				& 	\vspace{-0.25in} 	\\	 \cline{1-3} 

 \multicolumn{1}{r|}{source}			& 2012			&   2014		\\  \cline{1-3}
 \multicolumn{1}{r|}{combined statistics, $\chi$PT}			&  3.7				& 1.4		\\
 \multicolumn{1}{r|}{wrong spin}		&  3.2			& 	NA	\\  \cline{1-3}
 \multicolumn{1}{r|}{HQ discretization}			& 0.3				& included	 		\\  \cline{1-3}
 \multicolumn{1}{r|}{inputs}			& 0.7				& {included}	 		\\  \cline{1-3}
 \multicolumn{1}{r|}{renormalization}				& 0.5				& {0.5}	  		\\  \cline{1-3}
 \multicolumn{1}{r|}{finite volume}				& 0.5				&{0.5}	  		\\ \cline{1-3}
 \multicolumn{1}{r|}{total}				& 5 				& {1.6}       	\\ \cline{1-3} 
 								& 				& 				 \vspace{-0.195in}	\\ \cline{1-3}
\end{tabular}
\end{center}
\caption{Comparison of the error budgets for $\xi$ from Ref.~\cite{Bazavov:2012zs} (2012) with this analysis (2014).
Here, ``included'' means these errors are now included via Bayesian priors in the combined statistical, $\chi$PT error.}
\label{tab-xi-error}
\end{table}

There are errors of $O(\alpha_s^2)$ in our calculation since 
the renormalization coefficients are calculated in perturbation theory at one-loop order. 
The one-loop coefficients for the $B$-meson mixing operators are $O(1)$, and we therefore estimate the error as the average of 
$\alpha_s^2$ from all four lattice spacings. This yields the error shown in Table~\ref{tab-error}. 
We are currently investigating the effect of using the mostly nonperturbative renormalization method introduced 
in Ref.~\cite{Harada:2001fi} for heavy-light currents. 
Because the MILC ensembles have large spatial volumes with $M_\pi L \gtrsim 3.8$, we expect finite volume errors to be a subdominant source of
error, contributing at the 1\% level or less. We are currently in the process of including finite volume corrections in the chiral expansion. The estimates
shown in Tables~\ref{tab-error} and \ref{tab-xi-error} are from our decay constant analysis \cite{Bazavov:2011aa}. 

\section{Conclusions and outlook}

We present nearly final systematic error budgets for our analysis of the matrix elements $\langle \mathcal{O}_i\rangle$ and $\xi$.  
Tables~\ref{tab-error} and \ref{tab-xi-error} show comparisons of our current error budgets with our previous results. We find significant
improvement in all cases. For $\xi$ we expect a final error of $\lesssim 2\%$, more than a factor of two smaller than our previous result. This is 
only in
part due to the fact that Ref.~\cite{Bazavov:2012zs} used a much smaller subset of MILC ensembles. 
Another factor is that  with simultaneous fits to all three operators in our present analysis there is no "wrong spin" error anymore. 
Once our results are final, we also plan to 
combine those for the $\langle \mathcal{O}_i\rangle$ with the companion analysis of the $B_s$ and $B_d$ decay constants 
\cite{neil14} to obtain results for the corresponding bag parameters. 

\section*{Acknowledgements}

This work was supported by the U.S.\ Department of Energy, the National Science Foundation, the Universities
Research Association, the MINECO, Junta de Andaluc\'ia, the European Commission, the German Excellence
Initiative, the European Union Seventh Framework Programme, and the European Union's Marie Curie COFUND
program.
Computation for this work was carried out at the Argonne Leadership Computing Facility (ALCF), the National Center
for Atmospheric Research (UCAR), the National Center for Supercomputing Resources (NCSA), the National
Energy Resources Supercomputing Center (NERSC), the National Institute for Computational Sciences (NICS),
the Texas Advanced Computing Center (TACC), and the USQCD facilities at Fermilab, under grants from the NSF
and DOE.

\end{document}